\documentclass[useAMS,usenatbib]{mn2e}
\usepackage{graphicx,epsfig}

\def\culgora{Cul 1709-231}

\def\asec{\ifmmode ^{\prime\prime}\else$^{\prime\prime}$\fi}

\def\degs{\ifmmode ^{\circ}\else$^{\circ}$\fi}
\def\amin{\ifmmode ^{\prime}\else$^{\prime}$\fi}
\def\asec{\ifmmode ^{\prime\prime}\else$^{\prime\prime}$\fi}

\def\h{\hbox{$^{\rm h}$}}
\def\m{\hbox{$^{\rm m}$}}
\def\s{\hbox{$^{\rm s}$}}

\def\EE#1{\times 10^{#1}}

\def\cm{\mbox{\,cm}}

\def\cm3{\mbox{\,cm$^{-3}$}}

\def\ergs{\mbox{\,erg~s$^{-1}$}}
\def\ergcmcms{\mbox{\,erg~cm$^{-2}$~s$^{-1}$}}

\def\lsim{\!\!\!\phantom{\le}\smash{\buildrel{}\over
 {\lower2.5dd\hbox{$\buildrel{\lower2dd\hbox{$\displaystyle<$}}\over
                                 \sim$}}}\,\,}
\def\gsim{\!\!\!\phantom{\ge}\smash{\buildrel{}\over
{\lower2.5dd\hbox{$\buildrel{\lower2dd\hbox{$\displaystyle>$}}\over
                               \sim$}}}\,\,}

\title{The origin of the diffuse non-thermal X-ray and radio emission
  in the Ophiuchus cluster of galaxies}

\author[M.~A P\'erez-Torres et al.] 
{M.A.\ P\'erez-Torres$^1$\thanks{E-mail: torres@iaa.es}, 
   F.\ Zandanel$^1$,
   M.A.\ Guerrero$^1$,
   S. Pal$^2$, 
   S. Profumo$^3$, 
   F. Prada$^{1, 4}$, 
\newauthor
   F. Panessa$^5$ \\
$^1$Instituto de Astrof\'{\i}sica de Andaluc\'{\i}a, CSIC, Apdo.
Correos 3004, E-18080 Granada, Spain \\
$^2$National Centre for Radio Astrophysics, Tata Institute of Fundamental Research, Pune, 411007, India\\
$^3$Santa Cruz Institute for Particle Physics (SCIPP) and Department of Physics,
University of California, Santa Cruz CA 95064, USA \\
$^4$Visiting research physicist at the Santa Cruz Institute for Particle Physics (SCIPP),
University of California, Santa Cruz CA 95064, USA \\
$^5$Istituto di Astrofisica Spaziale e Fisica Cosmica  (INAF), 
00133 Roma, Italy
}

\date{Accepted  2009 April 7 .
      Received  2009 April 7;
      in original form 2008 December 17
     }

\pagerange{\pageref{firstpage}--\pageref{lastpage}} \pubyear{2005}
\def\LaTeX{L\kern-.36em\raise.3ex\hbox{a}\kern-.15em
T\kern-.1667em\lower.7ex\hbox{E}\kern-.125emX}

\begin{document}
\label{firstpage}
\maketitle


\begin{abstract}

We present high resolution 240 and 607 MHz GMRT radio observations,
complemented with 74 MHz archival VLA radio observations of the
Ophiuchus cluster of galaxies, whose radio mini-halo has been recently
detected at 1400 MHz. We also present archival \emph{Chandra} and
\emph{XMM-Newton} data of the Ophiuchus cluster.  Our observations do
not show significant radio emission from the mini-halo, hence we
present upper limits to the integrated, diffuse non-thermal radio
emission of the core of the Ophiuchus cluster.  The \emph{XMM-Newton}
observations can be well explained by a two-temperature thermal model
with temperatures of $\simeq$1.8 keV and $\simeq$9.0 keV,
respectively, which confirms previous results that suggest that the
innermost central region of the Ophiuchus cluster is a cooling core.  This
result is consistent with the occurrence of a mini-halo, as expected
to be found in hot clusters with cool cores.  We also used the
\emph{XMM-Newton} data to set up an upper limit to the (non-thermal)
X-ray emission from the cluster.  We also emphasize that the
non-thermal X-ray emission obtained with \emph{XMM-Newton} and
\emph{INTEGRAL} cannot be produced by the putative AGN of the galaxy
at the cluster center.

The combination of available radio and X-ray data has strong
implications for the currently proposed models of the spectral energy
distribution (SED) from the Ophiuchus cluster. In particular, a
synchrotron+IC model is in agreement with the currently available
data, if the average magnetic field is in the range
(0.02$-$0.3)$\mu$G. A pure WIMP annihilation scenario can in principle
reproduce both radio and X-ray emission, but at the expense of
postulating very large boost factors from dark matter substructures,
jointly with extremely low values of the average magnetic
field. Finally, a scenario where synchrotron and inverse Compton
emission arise from PeV electron-positron pairs (via interactions with
the CMB), can be ruled out, as it predicts a non-thermal soft X-ray
emission that largely exceeds the thermal Bremsstrahlung measured by
\emph{INTEGRAL}.

\end{abstract}



\section{Introduction}
\label{intro}

Clusters of galaxies are the largest bound structures of the Universe
and, according to hierarchical scenarios of structure formation, the
latest ones to form. They are filled by a hot ($10^7-10^8$ K) plasma,
called intra-cluster medium (ICM), and thus radiate in soft X-ray
bands through thermal Bremsstrahlung.  In hierarchical cosmological
scenarios, the most massive clusters form as a result of merging of
smaller clusters. The current scenario predict that a fraction of the
energy dissipated during the merging is channeled into particles
acceleration via shocks and turbulence (e.g. Sarazin 1999a; Blasi,
Gabici \& Brunetti 2007).  This lead to a complex population of
primary protons and electrons in the ICM.  Models predict that a large
population of \emph{secondary electrons} could be injected by
collisions between relativistic and thermal protons in the
ICM. Alternatively, the ICM relativistic electrons population could be
\emph{re-accelerated in situ} by various mechanisms associated with
turbulence in massive merger events. The recent observation of a radio
halo with a very steep synchrotron spectrum in Abell~521 by Brunetti
et al. (2008) supports the turbulence acceleration mechanism.  This
electron population is expected to be responsible for both the
synchrotron radio emission and the hard X-ray emission. The origin of
the latter emission is disputed. It has been proposed that it is due
to inverse-Compton (IC) scattering of relativistic electrons with the
cosmic microwave background (CMB) (Atoyan \& V\"olk 2000, Ensslin \&
Biermann 1998, Sarazin 1999b), or to a population of PeV electrons
that would radiate in hard X-rays through synchrotron emission
(Timokhin, Aharonian \& Neronov 2004, Inoue, Aharonian \& Sugiyama
2005). Those latter models have the disadvantage that do not explain
the well-known synchrotron radio emission.

Extended radio emission in clusters of galaxies has been known to
exist for a long time (see e.g. Feretti \& Giovannini 2007). However,
the detection of the putative hard X-ray emission has been rather
elusive, apart from some rather weak and controversial detections of a
hard tail in the X-ray spectrum of Coma (Fusco-Femiano et al. 2004)
and Abell 2256 (Fusco-Femiano et al. 2005) with \emph{BeppoSAX}, and
the \emph{Chandra} observations of the Perseus cluster (Sanders et
al. 2004, Sanders, Fabian \& Dunn 2005). Nevertheless, \emph{INTEGRAL}
failed to find an X-ray non-thermal component in the Coma cluster
(Renaud et al. 2006), and recent \emph{XMM-Newton} observations of
Perseus (Molendi \& Gastaldello 2008) did not confirm Sanders et
al. (2004) and Sanders, Fabian \& Dunn (2005) results.

The Ophiuchus cluster is a nearby (z=0.028, Johnston et al. 1981) rich
cluster located above the direction of the Galactic Center
($l=0.5^\circ, b=9.4^\circ$), with a very high plasma temperature
($kT\sim10$ keV).  Johnston et al. (1981) claimed, based on radio data
available at the time, that the cluster was associated with the
steep-spectrum radio source MSH 17-203 (also dubbed Cul 1709-231). The
identification of this radio source as a radio halo from the Ophiuchus
cluster would thus imply the presence of relativistic electrons, and
hence the presence of a non-thermal, high-energy tail in the X-ray
band would be expected. And, in fact, while this paper was being refereed,
Govoni et al. (2009) and Murgia et al. (2009) have reported
the detection of a radio mini-halo  in the Ophiuchus cluster (see Fig. 6 in
Govoni et al. 2009), using deep VLA 1400 MHz observations.

\begin{figure}
\begin{center}
\resizebox{8.6 cm}{!}{\includegraphics[width=8.5cm]{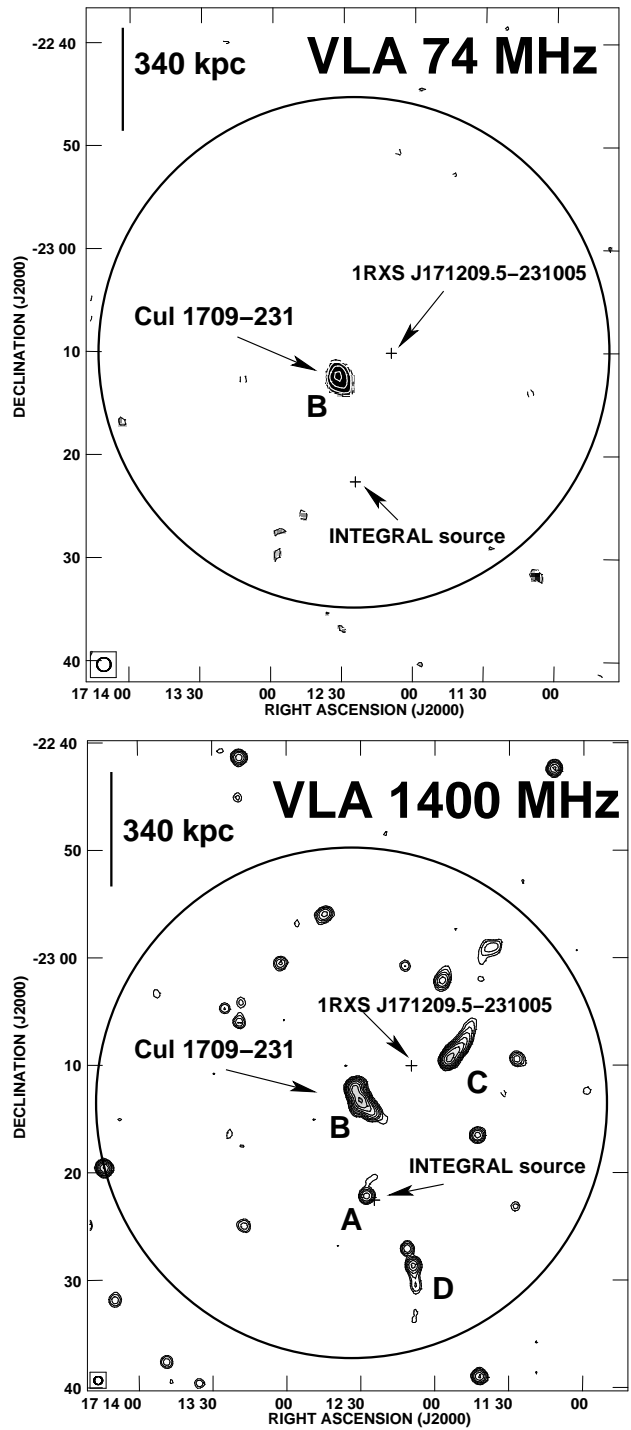}}
\caption[]{VLSS radio image at 74~MHz (top) and NVSS radio image at
  1400~MHz (bottom) of the Ophiuchus Cluster. In both panels the
  positions of the \emph{INTEGRAL} X-ray detection and of the nearest
  X-ray source 1RXS J171209.5-231005 are shown.  The synthesized beams
  (FWHM) at 74 MHz and 1400 MHz were (80''$\times$80'') and
  (45''$\times$45''), respectively, and are drawn at the bottom left
  of each panel.  The big circles in the panels correspond to the
  approximate size of the synthesized beam of the FLEURS and SYDNEY
  (big circle in top panel) and of the OSU (big circle in the bottom
  panel) radio observations (see Table \ref{all_data}), carried out in
  the 70's. Note that the radio emission of a large amount of
  different sources enters the beam, and therefore heavily
  overestimates the flux at those frequencies. The off-source rms
  noise in the VLSS and NVSS images is of 220 mJy/b and 0.4 mJy/b,
  respectively, and contours are drawn starting at three times the
  quoted rms, increasing in steps of $\sqrt{3}$.}
\label{fig,vla}
\end{center}
\end{figure}

Watanabe et al. (2001), based on \emph{ASCA} data, found that the core
of the cluster has an angular diameter of 6.4'.  Measurements of the
ICM temperature vary from 8.5$\pm$0.5~keV (INTEGRAL; Eckert et
al. 2008) up to 9.5$^{+1.4}_{-1.1}$~keV (Swift/BAT; Ajello et
al. 2008).  Watanabe et al. also found a large (20'$\times$30'), hot
($kT>13$ keV) region, 20' west of the cluster center, from which they
concluded that the cluster is not dynamically relaxed, and suggested
it experienced a major merging event in the recent past ($t \lsim 1$
Gyr).  Eckert et al. (2008) have recently reported a tentatively
resolved ($\sim$5') X-ray source at the cluster center, and claimed
the presence of a non-thermal tail. Eckert et al. interpret the
non-thermal hard X-ray emission as due to IC radiation from
relativistic electrons scattered off the CMB in the cluster
medium. The evidence of a non-thermal hard X-ray tail in the Ophiuchus
cluster from \emph{INTEGRAL} could be the first significant detection
of such non-thermal X-ray emission from a galaxy cluster.  More recent
Suzaku observations of the Ophiuchus cluster by Fujita et al. (2008)
have, however, failed to detect the non-thermal component claimed by
Eckert et al. (2008), although their quoted upper limit of
$2.8\times10^{-11}$~erg~cm$^{-2}$~s$^{-1}$ in the $20-60$~keV energy
band is still compatible with the \emph{INTEGRAL} detection.  Fujita
et al. (2008) found that the cluster was hot, except for the innermost
$\sim$50 kpc, and that the intra-cluster medium (ICM) was in
ionization equilibrium state. From those results, Fujita et
al. conclude that Ophiuchus cluster is not a major merger, but one of
the hottest clusters with a cool core.

Even more recently, Ajello et al. (2008) have found, combining Swift/BAT
and \emph{Chandra} spectra, an upper limit on the Ophiuchus
non-thermal X-ray emission in the $20-60$~keV band, of
$7.2\times10^{-12}$~erg~cm$^{-2}$~s$^{-1}$, which is below the
\emph{INTEGRAL} detection.  Nevertheless, the claimed \emph{INTEGRAL}
detection, the Suzaku and the combined Swift/BAT and \emph{Chandra}
upper limits are consistent.

In this Paper, we present and discuss archival and new radio data for
the Ophiuchus cluster, along with archival X-ray data.  Together with
the published X-ray data of the cluster, we model all the data to shed
light on the origin of the diffuse non-thermal emission from this
cluster of galaxies.  In Section 2, we present the VLA and GMRT radio
data from the Ophiuchus cluster, along with archival \emph{Chandra}
and \emph{XMM-Newton} data. In Section 3, we model the SED of the
Ophiuchus cluster using several different scenarios.  Finally, we
summarize our main results and their implications in Section 4.  All
sky positions in the paper are for the J2000.0 equinox. We assume
cosmological parameters of $\Omega_{0}=0.3$, $\lambda_{0}=0.7$ and
$H_{0}=70$~km~s$^{-1}$~Mpc$^{-1}$. For these parameters, our assumed
redshift of 0.028 to the Ophiuchus cluster (Johnston et al. 1981)
corresponds to a luminosity distance of 122.6~Mpc and a comoving
distance of 126.03~Mpc. At this distance, 1'' corresponds to 560 pc.


\section{Observations and results}
\label{observations}

We present in Table~\ref{all_data} a summary of all the relevant data
for the Ophiuchus cluster, spanning 11 decades of energy from the
radio up to gamma-rays, and which include both data from the
literature and obtained by us (this paper). We present the main
esults of these observations in Figures 1 through 4, where we 
show high-resolution radio and X-ray images of the Ophiuchus cluster of
galaxies.

\subsection{Radio observations}
\label{radio}

\subsubsection{Data from the literature}

The first five rows of Table~\ref{all_data} show the radio
measurements of the Ophiuchus cluster of galaxies, compiled by
Johnston et al. (1981), and have been used in the past to model the
SED from the Ophiuchus cluster (e.g., Eckert et al. 2008, Profumo et
al. 2008).  We note that these rather old radio data have
significantly poorer resolution than is currently available with
existing radio interferometer arrays, and therefore the use of higher
resolution radio data might have a relevant impact on the modelling of
the spectral energy distribution (SED) of the Ophiuchus cluster.  In
particular, the old radio data at 30, 86 and 1400~MHz cover a too
large area of sky (see the drawn circles in Figure \ref{fig,vla}), so
that the radio emission of a large amount of different sources enters
into the beam, and therefore heavily overestimates the flux at those
frequencies.

\subsubsection{VLA data}
\label{vla}

Fig. \ref{fig,vla} shows publicly archival Very Large Array (VLA)
radio images of the Ophiuchus Cluster at 74~MHz (top), taken as part
of the VLA Low-frequency Sky Survey (VLSS; Cohen et al. 2007), and at
1400~MHz (bottom; NRAO VLA Sky Survey - NVSS; Condon et al. 1998).
Note that the two Culgora data measurements at 80 and 160~MHz in Table
\ref{all_data} were obtained with resolutions of a few arcmin. The
dominant radio source appears to be Cul 1709-231 (RA=17\h12\m29.70\s,
DEC=-23\degs 12\amin 6\asec). This source position coincides within
the errors with the only radio source that is clearly detected at 74
MHz in the VLSS image (upper panel of Figure \ref{fig,vla}), which has
a resolution of 80''. The emission peak ($\approx 4.2$ Jy/beam) does
indeed correspond to the source Cul 1709-231, for which we obtain
RA=17\h12\m31.30\s and DEC=$-23$\degs12\amin24.26\asec, with a
positional uncertainty of 2.6'' in each coordinate.  In the 74 MHz
VLSS image (top panel of Figure \ref{fig,vla}), Cul 1709-231 is
extended, and has a total flux of $\sim$8.0 Jy. However, its position
is $\sim$17' off the maximum of emission obtained with \emph{INTEGRAL}
by Eckert et al. (2008).  (We used Fig. 1 in Eckert et al. to
determine the position of the maximum of emission of the
\emph{INTEGRAL} source: RA=17\h12\m24.50\s,
DEC=$-23$\degs22\amin35.0\asec, with an estimated uncertainty of
$\approx 1.8'$ in each coordinate).  Since the core radius is 3.2'
(Watanabe et al. 2001, Eckert et al. 2008), it is unlikely that Cul
1709-231 can be physically related with the core of the cluster. In
fact, the very recent discovery of the radio mini-halo close to source
A shows that \culgora\ is not related at all with the cluster core.

The archival 1400~MHz NVSS image (bottom panel of Figure
\ref{fig,vla}) has two times better resolution than the 74 MHz image,
and shows a considerably larger number of radio sources. This comes as
no surprise, since the off-source rms in the VLSS image of the
Ophiuchus region is 220 mJy/b, while that of the NVSS image is
$\approx$0.4 mJy/b, i.e., the NVSS image is 550 times more sensitive,
and therefore is able to detect many sources that went undetected in
the VLSS image.  In particular, we found a relatively faint ($S_\nu
\approx 50$ mJy) radio source (dubbed A in the bottom panel of
Fig. \ref{fig,vla}) at RA=17\h12\m27.47\s,
DEC=$-23$\degs22\amin04.9\asec. This position is $\sim$6'' away from
the optical position of an elliptical, cD galaxy at $z=0.028456$
(2MASX J17122774-2322108), which may suggest that the radio emission
arises from it (see also sections \ref{gmrt}, \ref{x-rays}
and Figure \ref{fig,chandra}). The off-source rms around this source
is of about 1.0 mJy/b.

Very recently, Govoni et al. (2009) and Murgia et al. (2009) have
detected a (radio) mini-halo in the Ophiuchus cluster (see Fig. 6 in
Govoni et al. 2009), using deep VLA 1400 MHz observations. Murgia et
al. (2009) report a flux density of $\approx$106 mJy for this
mini-halo, and a size of $\approx 9 \times 12$ arcmin$^2$. It is also
not surprising that the NVSS archival image did not show evidence for
the mini-halo, since the off-source rms attained by Govoni et
al. (2009) was of 0.1 mJy/b, five times deeper than the overal
off-source rms in the NVSS image we used (and ten times deeper than
the off-source rms around the radio emission detected from the cD
galaxy - see previous paragraph - which is the relevant figure). Thus,
while both our archival NVSS image and that of Govoni et al. clearly detect the
central, compact emitting source, the much poorer sensitivity of the
public NVSS image prevented us from detecting the mini-radio halo at
1400 MHz.

\subsubsection{GMRT radio observations}
\label{gmrt}

We observed the field of the Ophiuchus Cluster on 23 May 2008 for
about two hours, using the Giant Metrewave Radio Telescope (GMRT) near
Pune, India.  The GMRT is an array of thirty fully steerable,
parabolic radio antennas of 45 meter diameter each, spread over an
area of 25 km in size. 16 of the antennas are displayed along three
arms, with nearly a {\it `Y'} shape, and the remaining 14 antennas are
placed within one kilometer of the central part of the array.  We
observed simultaneously at the sky frequencies of 240 and 607 MHz,
using synthesized bandwidths of 6 and 32 MHz, respectively.  We
observed in spectral line mode, with channels widths of 125 KHz, to be
able to accurately remove the radio frequency interference (RFI),
which affects heavily low-frequency radio observations.  We used
3C~286 as flux and band-pass calibrator, and used Baars et al. (1977)
to set the flux density scale, and used B1822-096 as phase calibrator.
We edited, flagged, and calibrated the data within the NRAO
Astronomical Imaging Processing Software (\emph{AIPS}).  After these
calibration steps, we averaged our data to increase the signal to
noise ratio, resulting in effective channel widths of 1 and 2 MHz at
240 and 607 MHz, respectively. Because of the small bandwidth, sources
in our fields are not affected by bandwidth smearing problems.  We
then imaged and selfcalibrated the data (see Fig. \ref{fig,gmrt}) also
within \emph{AIPS}, and corrected the images for their beamshape,
yielding final resolutions of $25.8'' \times 13.5''$ and $7.1'' \times
6.1''$ at 240 and 607 MHz, respectively.

\begin{figure*}
\begin{center}
\resizebox{17.2 cm}{!}{\includegraphics[width=17.0cm]{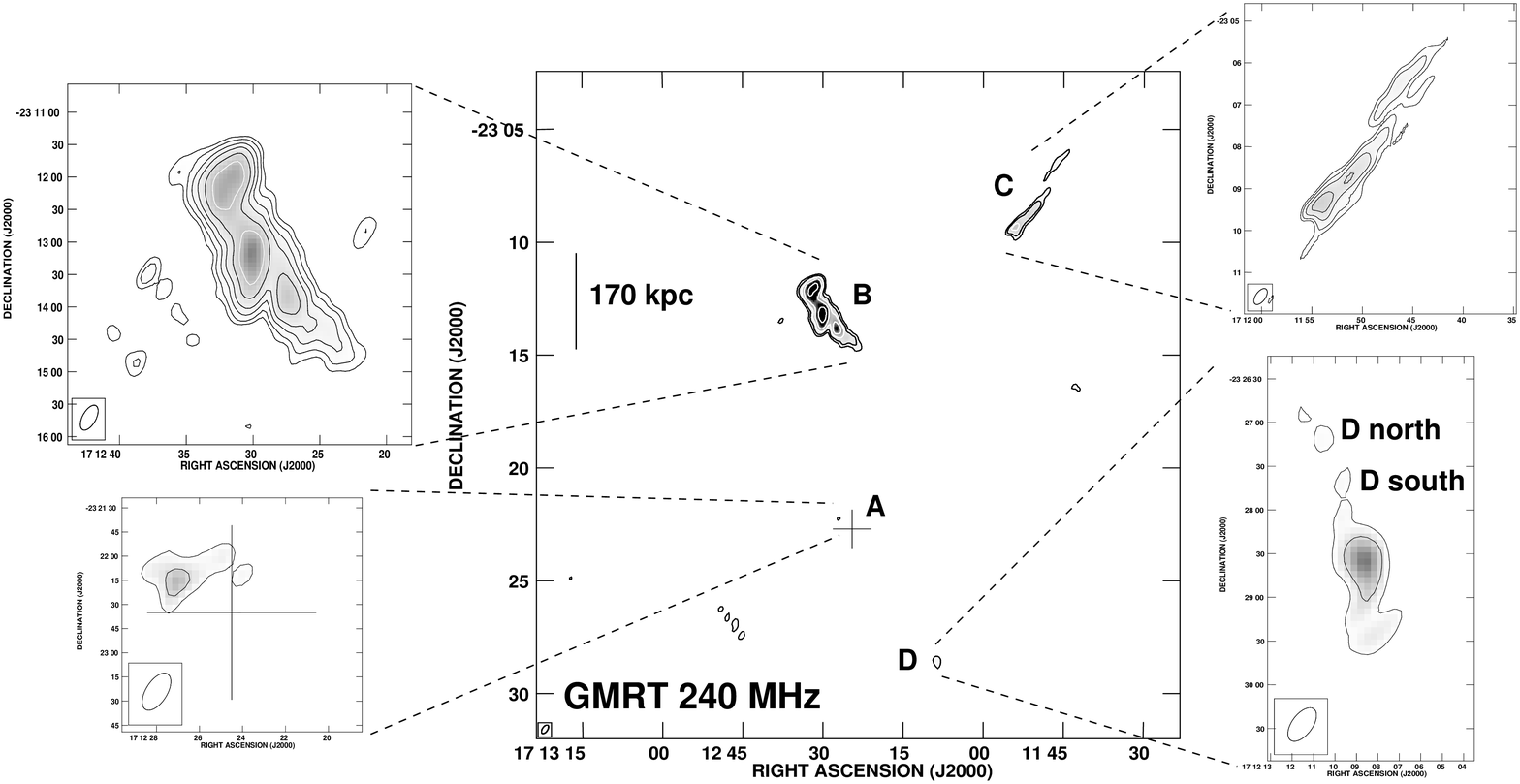}}
\resizebox{17.2 cm}{!}{\includegraphics[width=17.0cm]{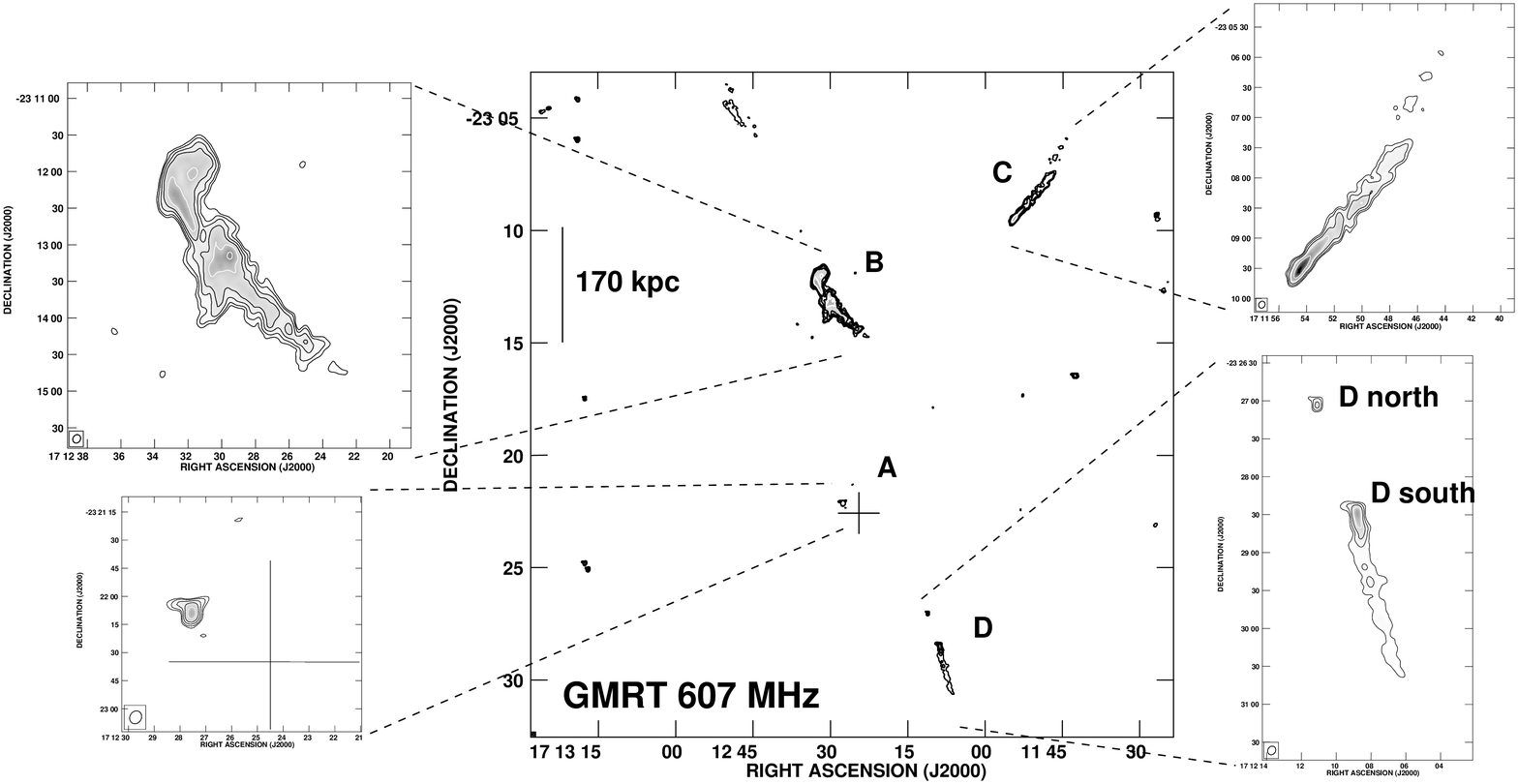}}
\caption[]{ GMRT images at 240 MHz (top) and 607 MHz (bottom) of the
  Ophiuchus Cluster region.  In both panels, the cross marks the
  position of the \emph{INTEGRAL} X-ray detection.  The synthesized
  beams (FWHM) at 240 MHz and 607 MHz were $26.1'' \times 13.5''$ and
  $7.1'' \times 6.1''$, respectively, and are drawn in the bottom left
  part of each panel (the linear size shown in the images have been
  calculated for an assumed redshift of $z = 0.028$).
  The  off-source rms noise  of the 240 and 607 MHz images is of 5.0 mJy/b and
  0.4 mJy/b, respectively. Contours are drawn starting at three
  times the quoted rms, and increase in steps of $\sqrt{3}$.
}
\label{fig,gmrt}
\end{center}
\end{figure*}

\begin{table*}
\caption[t]{Summary of the available radio, X and gamma-ray non-thermal measurements of the Ophiuchus cluster.}
\begin{center}
\begin{tabular}{lllllll}
\hline
Instruments &  $\nu$  &  E  &  FWHM  &  F  &   EF  &  References \\
          &  [MHz] &  [eV] & [']  &  [Jy] & [erg~cm$^{-2}$~s$^{-1}$]  & \\
\hline
FLEURS   &  $29.9$   &  $1.24\times 10^{-7}$  &  48$\times$48          &  $215\pm16$   &  $(6.43\pm0.48)\times 10^{-14}$  &  [1]  \\
CULGORA  &  $80$     &  $3.31\times 10^{-7}$  &  3.7$\times$3.7         &  $23\pm2$     &  $(1.84\pm0.16)\times 10^{-14}$  &  [1]  \\
SYDNEY   &  $86$     &  $3.56\times 10^{-7}$  &  50$\times$50          &  $42\pm4$     &  $(3.61\pm0.34)\times 10^{-14}$  &  [1]  \\
CULGORA  &  $160$    &  $6.62\times 10^{-7}$  &  1.9$\times$1.9         &  $6.4\pm0.1$  &  $(1.02\pm0.02)\times 10^{-14}$  &  [1]  \\
OSU      &  $1415$   &  $5.85\times 10^{-6}$  &  40$\times$40  &  $0.8\pm0.2$  &  $(1.13\pm0.28)\times 10^{-14}$  &  [1]  \\
\hline
VLA   &  $74$    &  $3.06\times 10^{-7}$  &  1.33$\times$1.33  &  $<13.43$   &  $<9.89 \times 10^{-15}$  & This work \\
GMRT  &  $240$   &  $9.93\times 10^{-7}$  &  0.44$\times$0.23  &  $<5.34$    &  $<1.32 \times 10^{-14}$  & This work \\
GMRT  &  $607$   &  $2.51\times 10^{-6}$  &  0.12$\times$0.10  &  $<3.60$    &  $<2.24 \times 10^{-14}$  & This work \\
VLA   &  $1400$  &  $5.79\times 10^{-6}$  &  1.52$\times$0.67  &  $0.106\pm0.030$  &  $ (1.48\pm0.42) \times 10^{-15}$  & [2] \\
\hline
XMM-Newton (EPIC)  &  $1.2 - 24.2 \times 10^{11}$  &  $0.3 - 9 \times 10^{3}$ &    0.1   &  $\lsim 1.14$  &  $\lsim 1.44 \times 10^{-11}$   & This work \\ 
\hline
SUZAKU (90\% UL)    &  $0.5 - 1.5 \times 10^{13}$   &  $2 - 6 \times 10^{4}$    &    2 &  $<2.9\times 10^{-7}$   &  $<2.8\times 10^{-11}$    &  [3] \\
\hline
Swift/BAT (90\% UL) &  $0.5 - 1.5 \times 10^{13}$   &  $2 - 6 \times 10^{4}$    &    20   &  $<7.6\times 10^{-8}$   &  $<7.2\times 10^{-12}$    &  [4] \\
\hline
INTEGRAL (IBIS/ISGRI)   &  $1.21\times 10^{13}$  &  $(4-6)\times 10^{4}$  &  12 &  $(7.93\pm1.32)\times 10^{-8} $  &  $(9.61\pm1.61)\times 10^{-12}$  & [5] \\
INTEGRAL (IBIS/ISGRI)  &  $1.69\times 10^{13}$  &  $(6-8)\times 10^{4}$  &  12  &  $(4.97\pm2.66)\times 10^{-8} $  &  $(8.41\pm4.49)\times 10^{-12}$  & [5] \\
\hline
EGRET  (2$\sigma$ UL) & $2.4\times 10^{17} - 10^{19}$  &  $10^{8} - 10^{10}$     &   (6-30)   &  $<3.8\times 10^{-13}$  &  $<9.1\times10^{-12}$   &  [6] \\
\hline
\end{tabular}
\end{center}
\label{all_data}
\begin{flushleft}
 Note that the two CULGORA flux density values correspond to
 measurements of the cluster Source B, since its resolution was good
 enough.  On the contrary, the FLEURS, SYDNEY and OSU had poorer
 resolutions, thus covering a too large area of sky (see Figure
 \ref{fig,vla} and text for details), and therefore including the
 radio emission from multiple sources in the field.
\\ References: [1]
 Johnston et al. (1981), [2] Murgia et al. (2009), [3] Fujita et al. (2008), [3] Ajello et al. (2008),  
 [5] Eckert et al. (2008), [6] Reimer at al. (2003).
\end{flushleft}
\end{table*}

\begin{table*}
\caption[t]{List of the sources found in the GMRT images field of view with the corresponding NED coincident sources.}
\begin{center}
\begin{tabular}{lllllllll}
\hline
Source & $\nu$ & RA  &  DEC  &  Peak  &  Total Flux  &  Coincidence & $z$ & Type \\
       & [MHz] &     &       &  [mJy/beam]  & [mJy]  &               &     & \\
\hline
A      &  $607$  &  17\h 12\m 27.56\s  &  -23\degs 22\amin 09.5\asec  &  
  $29.1$  &  $37.8\pm1.0$  &  2MASX~J17122774-2322108  & 0.028 &  G-cD \\
       &  $240$  &                            &                                 &  $33$    &  $38\pm5$      &                       &    & \\
\hline
B      &  $607$  &  17\h 12\m 29.51\s  &  -23\degs 13\amin 09.4\asec  &  $38.7$  &  $2020\pm21$   &  Cul~1709-231        &  -   & R-S \\
       &         &                            &                                 &          &                &  NVSS~J171224-231424   &  -  & R-S  \\
       &  $240$  &                            &                                 &  $391$   &  $4853\pm30$   &                     &      & \\
\hline
C      &  $607$  &  17\h 11\m 54.52\s  &  -23\degs 09\amin 32.8\asec  &  $31.3$  &  $579\pm7$     &  2MASX~J17115542-2309423  &  0.027 &G-E\\
       &         &                            &                                 &          &                &  PMN~J1711-2308       &  -  & R-S \\
       &  $240$  &                            &                                 &  $102$   &  $1117\pm7$    &                        &   & \\
\hline
D (north) &  $607$  &  17\h 12\m 11.10\s  &  -23\degs 27\amin 03.9\asec  &  $8.9$   &  $11.2\pm0.8$  &  NVSS~J171211-232705    & -  & R-S\\
          &  $240$  &                            &                                 &  $17$    &  $83\pm5$      &                           & \\
\hline
D (south) &  $607$  &  17\h 12\m 08.769\s  &  -23\degs 28\amin 29.62\asec  &  $20$    &  $161\pm2$     &  2MASX~J17120908-2328263  & 0.025 & G-E\\
          &         &                            &                                 &          &                &  NVSS~J171208-232839    &  -  & R-S  \\
          &  $240$  &                           &                                 &  $44$    &  $263\pm5$     &                         &  & \\
\hline
\end{tabular}
\end{center}
\label{radio_sources}
\begin{flushleft}
Source positions (J2000.0) corresponding to the local maxima of our 607 MHz GMRT
image. The type, classification, and redshift $z$ for the sources are
obtained from NED (Type: G-cD, G-E and R-S stand for elliptical cD
galaxy, elliptical galaxy and radio source, respectively).
\end{flushleft}
\end{table*}

\subsubsection{Sources in the field of the Ophiuchus cluster and upper limits to its diffuse radio emission}
\label{sources}

We performed an extensive search for positional coincidences of the
local maxima in our 607 MHz GMRT image, using the NASA Extragalactic
Database (NED), aimed at confirming (or ruling out) the physical
belonging of the radio sources to the Ophiuchus cluster of galaxies.
The search for counterparts was performed within a 3\amin radius from
the given positions and we summarize our results in
Table~\ref{radio_sources}.  The position of the radio and X-ray peak
(\emph{Chandra}, see section \ref{x-rays}) of source A lies very close to 
the position of 2MASX J1722774-2322108 (Hasegawa et al. 2000), an
elliptical cD galaxy at $z$=0.028, and which is at the center of the
Ophiuchus cluster.  
Source B coincides with source Cul 1709-231
and with NVSS J171224-231424, which is approximately located at
1.5$\amin$ in the SE direction from the local maximum of radio
emission. NED classifies those sources simply as radio sources where
Cul 1709-231 is described as extended; no redshift measurements are
available for those sources.  The source C coincides with the source
2MASX J17115542-239423, an elliptical galaxy with redshift $z =
0.027$, and with PMN J1711-2308, located at $\sim 1\amin$ in the NE
direction, from the source local maximum and it is classified as a
radio source with no redshift measurement available.  Source D-north
perfectly coincides with NVSS J171211-232705, a radio source with no
redshift measurement available.  Finally, the source D-south position
matches extremely well with the position of 2MASX J17120908-2328263,
an elliptical galaxy with redshift z = 0.025, and with the source NVSS
J171208-232839, classified as a radio source and with no redshift
measurement available.

We therefore consider that source A is at the center of the Ophiuchus
cluster, and suggest that sources C and D-south are very likely
members of the Ophiuchus cluster, given their redshift. Visual
inspection of archival NED optical images reveals that PMN
J171224-231424 and NVSS J171211-232705 are probably galaxies. However,
spectroscopic follow-up is needed in order to determinate their
redshift and confirm their cluster membership. This is very important
above all for the Cul 1709-231, NVSS J171224-231424 and NVSS
J171208-232839 sources, for which the image inspection reveals no
clear evidence for optical counterparts. In particular, the Culgora
source characterization is very important in order to establish its
properties and cluster membership, given that, in the past, it was
confused with the central elliptical cD galaxy (source A) due to the
poor resolution of the old radio data (Johnston et al. 1981).

We also used our 240 and 607 MHz GMRT images (see
Fig. \ref{fig,gmrt}), along with the archival 74 VLA image (see
Fig. \ref{fig,vla}), to obtain upper limits to the diffuse,
non-thermal radio emission from the Ophiuchus cluster of galaxies.  In
particular, we used an area of $\approx 9 \times 12$ arcmin$^2$, very
similar to the size of the radio mini-halo found by Govoni et
al. (2009), and centered in source A (the core of the cluster). We
then excised the flux density of this point-like source in our images,
and present upper limits to those measurements in Table
\ref{all_data}. These were obtained by multiplying the angular area
above (after being converted into a number of beams at each
frequency), by the off-source rms 74,240, and 607 MHz (220, 5, and 0.4
mJy/b, respectively).  From the upper limits at low frequencies, and
the detection of the Ophiuchus cluster (mini-halo) at 1400 MHz, we
obtain an upper limit to the (radio) spectral index of $\beta \lsim
1.73$ ($S_\nu \propto \nu^{\beta}$). If we adopt a 3~$\sigma$ upper
limit to the non-detected extended radio emission, this yields an
upper limit for the spectral index of $\beta \lsim 2.09$.  Those
values are consistent with typical radio spectral indices found in
radio mini-halos ($\beta \gsim 1.5$; Ferrari et al. 2008).


\subsection{X-ray observations}
\label{x-rays}

\begin{figure}
\begin{center}
\resizebox{8.6cm}{!}{\includegraphics[width=8.5cm]{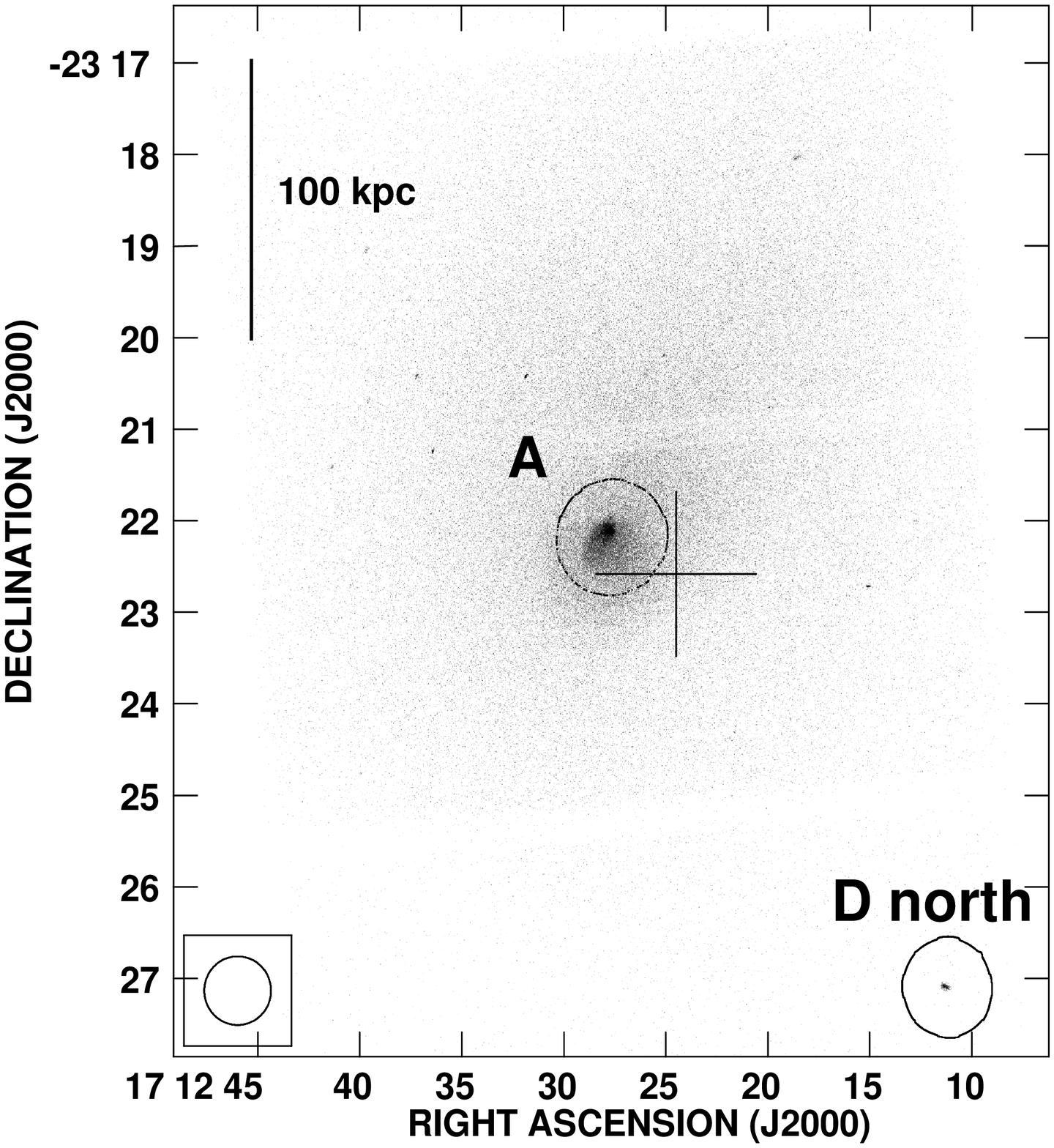}}
\resizebox{8.6
cm}{!}{\includegraphics[width=8.5cm,angle=0]{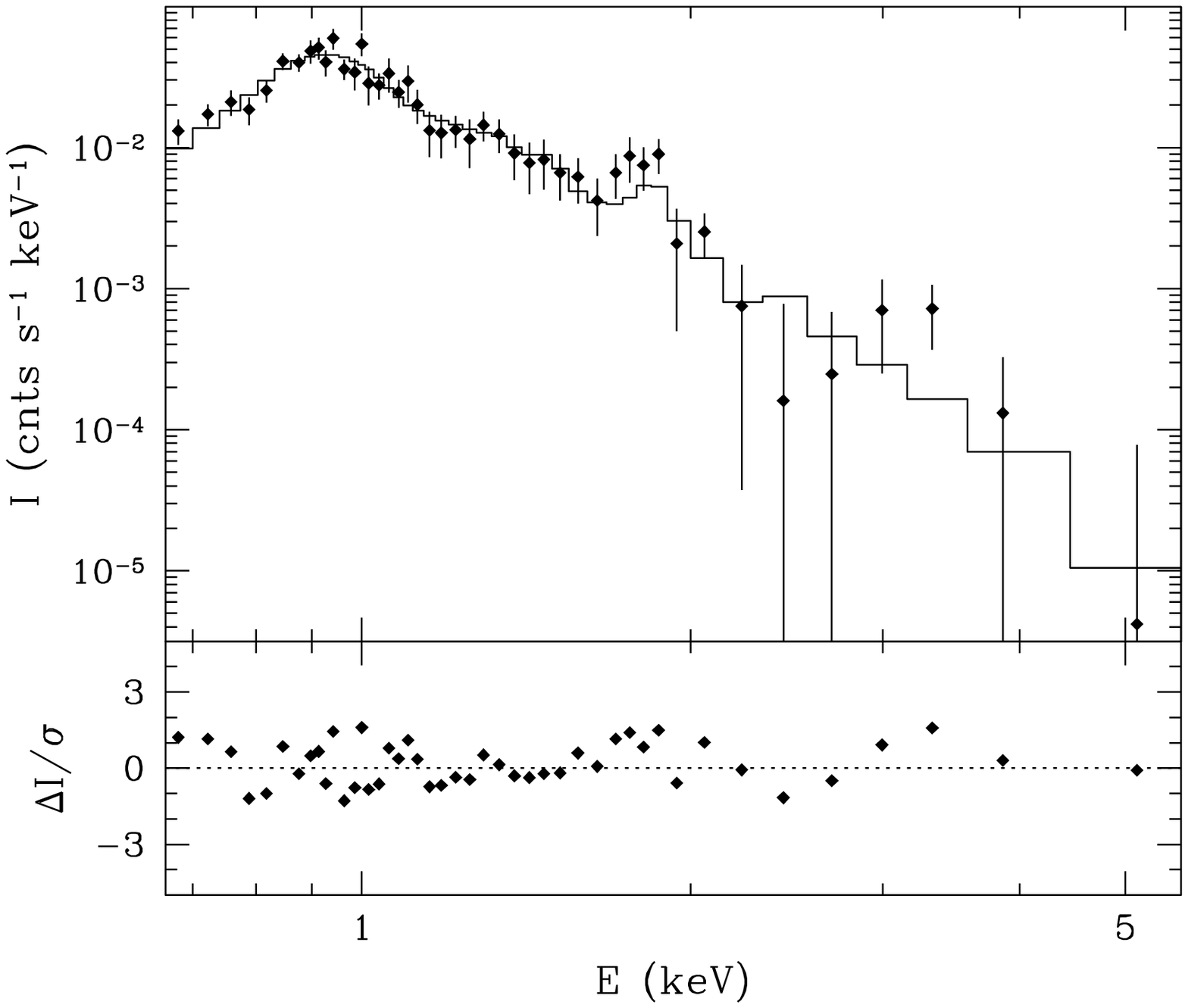}}
\caption[]{ {\it (Top)} \emph{Chandra} ACIS greyscale image of the
 core of the Ophiuchus cluster overplotted with 10 $\sigma$ NVSS 1400
 MHz radio contours.  The locations of sources A and D north,
 coincident with two point-like X-ray sources, are labeled on the
 plot.  The cross marks the position of the \emph{INTEGRAL} source
 detected by Eckert et al.\ (2008).  {\it (Bottom)}
 Background-subtracted \emph{Chandra} ACIS spectrum of the point
 source at the core of the Ophiuchus cluster overplotted with the
 best-fit MEKAL model.  The residuals of the best-fit model are shown
 in the lower panel.  }
\label{fig,chandra}
\end{center}
\end{figure}

The \emph{Chandra} X-ray Observatory observed the Ophiuchus cluster
for 51.2 ks (obs.\ ID 3200) on 2002 October 21.
The Advanced CCD Imaging Spectrometer (ACIS) instrument was used and
the core of the cluster was imaged on the back-illuminated CCD S3.
The data were processed using the standard data processing version
DS~7.6.9 and subsequently analyzed using the \emph{Chandra} Interactive
Analysis of Observations (CIAO) software package version 4.0.
The observations were not affected by any period of high-background,
and no time intervals had to be excised.
The final exposure time was 50.5 ks.

\emph{XMM-Newton} also observed the Ophiuchus cluster for 36.7 ks
(obs.\ ID 0505150101) during revolution 1416 on 2007 September 2.  The
European Photon Imaging Camera (EPIC) pn was used in the extended full
frame mode for 26.8 ks with the medium filter.  The data were
processed starting from the odf files with the \emph{XMM-Newton} SAS
software (version 7.1.2).  X-ray events corresponding to patterns
0-4 were selected.
Exposure has been filtered for periods of high background with an
effective exposure of 20 ks.
\emph{Chandra} and \emph{XMM-Newton} spectra
have been analyzed using HEASARC XSPEC v11.0.1 routines (Arnaud 1996).

We have used the \emph{Chandra} observations of the Ophiuchus cluster
to obtain a high-resolution X-ray image of its central region
(Figure~\ref{fig,chandra}-{\it top}). The \emph{Chandra} observations
are especially suited for the study of point-like sources, while the
extended, diffuse emission of this cluster is difficult to study
because its spatial extension is much larger than \emph{Chandra} ACIS
field of view.  Therefore, we did not use the \emph{Chandra}
measurements to estimate a limit to the diffuse X-ray emission of the
cluster, and hence this information is not included in Table
\ref{all_data}.  The analysis of the brightness radial profile of the
central region shows the presence of a relatively bright point-like
source at the core of the Ophiuchus cluster with position
RA=17\h12\m27.64\s, DEC=$-23^\circ$22'07".5 (the uncertainty in each
coordinate is $\approx$1''). The central source profile is clearly
point-like.  and cannot be modeled by classical cluster profile models
(such as King, Beta-models, etc.).  The location of this point source
is only 3.5'' away from the elliptical cD galaxy
2MASX\,J17122774-2322108.  Furthermore, Fig.~\ref{fig,chandra}-{\it
  top} shows that the NVSS 1400 MHz radio peak of the radio emitting
source A matches well with the location of the X-ray point source at
the core of the Ophiuchus cluster. This spatial coincidence suggests
that the radio emission from source A and the X-ray emission from the
point source at the core of the Ophiuchus cluster are associated.  The
radio emission from source A (see Table 2) indicates a flat spectrum, which
suggests that its radio emission is AGN-like, and thus the nuclear
activity of the cD galaxy 2MASX\,J17122774-2322108 can be claimed
responsible for the radio and X-ray emission.

The \emph{Chandra} ACIS X-ray spectrum of the point source at the core
of the Ophiuchus cluster was extracted using CIAO from a circular
region of radius 2$\arcsec$ centered at its position.  The
background-subtracted X-ray spectrum presented in the lower panel of
Fig.~\ref{fig,chandra} shows a bright peak at $\sim$0.9 keV, and
relatively fainter peaks at $\sim$1.5 and $\sim$1.8 keV.  This
spectral shape suggests the presence of line emissions as can be
expected from an optically-thin emitting plasma.  Indeed, the spectrum
is well fitted using a single temperature MEKAL model with plasma
abundances 0.3 $Z_\odot$ adequate for the central regions of the
Ophiuchus cluster (Watanabe et al.\ 2001), with a temperature $kT$ of
0.78 keV, and a column density of $4.2\EE{21}$ cm$^{-2}$.
The quality of the fit suggests that a non-thermal X-ray contribution
is negligible.
The unabsorbed flux in the 0.3 to 9.0 keV energy band is $F \approx
1.9\EE{-13}$\ergcmcms.  At the distance of the Ophiuchus cluster, this
implies an intrinsic luminosity $L_X \approx 3.3\EE{41}$\ergs. The
absence of a non-thermal X-ray component suggests that the central
nucleus is very weak and/or highly absorbed. Even if the Chandra X-ray
luminosity is entirely associated with the central active nucleus,
this must be of low luminosity.  Indeed, no recent strong AGN activity
has been observed at the cluster center (Dunn \& Fabian 2006).
Therefore, the activity of the nucleus of this galaxy cannot be
claimed responsible of the hard X-ray emission detected by
\emph{INTEGRAL}.

The \emph{XMM-Newton} EPIC-pn observations have been used to
investigate the X-ray emission from the core of the Ophiuchus cluster
given its sensitivity to low surface brightness diffuse X-ray
emission.  In order to asssess the occurrence of non-thermal emission
from the mini-halo at the core of the Ophiuchus cluster, a spectrum
has been extracted from a circular region of radius 5$\arcmin$
centered at the position of the cluster core that encompasses the
radio mini-halo region (Govoni, priv.\ comm.).  The corresponding
background spectrum has been obtained from a 50$\arcsec$ radius region
off-set $\sim$10$\arcmin$ from the cluster core in the same
CCD\footnote{ For comparison, we also considered two different
  background spectra: one extracted from the same observations from a
  distant circular region in EPIC-pn CCD 12 off-set $\sim$14.5$\prime$
  from the cluster core, and another one built composed of the sky
  background given by the \emph{ROSAT All Sky Survey} (RASS) and the
  particle background given by the \emph{XMM-Newton} closed filter
  data, kinldy provided by J.\ H.\ Nevalainen.  No significant
  differences were found in the background subtracted spectrum of the
  core of the Ophiuchus cluster using these three different background
  spectra, basically because the cluster brightness is $\gsim$100
  times brighter than the background emission.}.  The most updated
calibration files available at the time of the reduction were used to
generate the ancillary and detector response matrices using the
\emph{XMM-Newton} SAS \emph{arfgen} and \emph{rmfgen} tasks.

The background-subtracted spectrum displayed in Figure~\ref{fig,xmm}
shows a bright and broad peak at $\sim$1.08 keV, and fainter peaks at
$\sim$1.4 keV, and $\sim$1.7 keV.  It is also very noticeable a hard
energy tail up to 10 keV, as well as a prominent double-peaked
emission line at $\sim$6.4 keV and $\sim$6.7 keV, which corresponds to
Fe-K emission lines.  The spectrum can be relatively well fitted
($\chi^2$/DoF=1.16) by a two-temperature MEKAL model.  The best-fit
model has a hydrogen column density $N_{\rm H}=3.2\times10^{21}$
cm$^{-2}$, and temperatures of $kT=1.8\pm0.3$ keV and $kT=9.0\pm0.2$
keV.  The total unabsorbed flux in the 0.3 to 9.0 keV energy band is
$F \approx 2.5\EE{-10}$\ergcmcms, corresponding only a flux of $F
\approx 4.5\EE{-12}$\ergcmcms to the soft component, i.e., the X-ray
spectrum of this region is dominated by the hard emission associated
to the outermost regions of the cluster seen in projection towards its
core.
This result is consistent with the conclusion obtained by Fujita et al. (2009)
that Ophiuchus hosts a cool core, and  yields support to their suggestion
that Ophiuchus is not the result of a major merger event.

The addition of an additional power law component with an assumed photon
index of $\Gamma=1.6$ (Eckert et al.\ 2007) does not improve significantly
the fit ($\chi^2$/DoF=1.17), and thus such a component can be discarded.
The 3-$\sigma$ upper limit of the unabsorbed flux of such component has
been estimated to be  $F \le 1.44\EE{-11}$\ergcmcms.
At the distance of the Ophiuchus cluster, this implies an upper limit
to the intrinsic luminosity of non-thermal emission of $L_X(pow) \le
2.5\EE{43}$\ergs.

%

\begin{figure}
\begin{center}
\resizebox{8.6cm}{!}{\includegraphics[width=8.5cm]{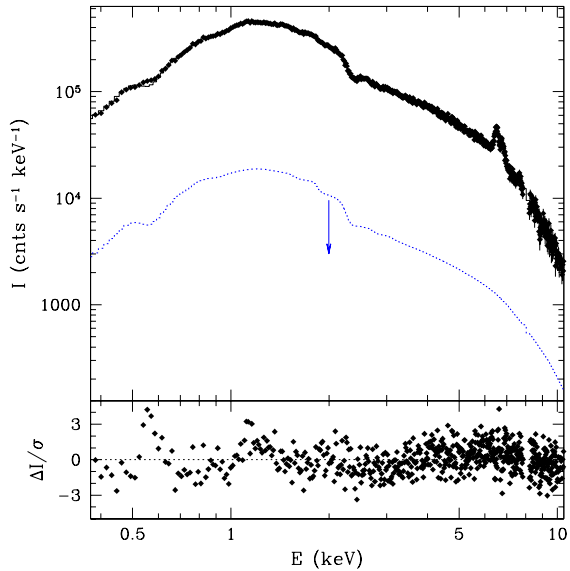}}
\caption[]{
Background-subtracted \emph{XMM-Newton} EPIC-pn spectrum of the central
region of the Ophiuchus cluster overplotted with the best-fit
two-temperature MEKAL model.
The 3-$\sigma$ upper limit for a non-thermal component is shown as
a dotted blue histogram.
The residuals of the best-fit two-temperature MEKAL model are shown
in the lower panel.
}
\label{fig,xmm}
\end{center}
\end{figure}


\section{Modelling the SED of the Ophiuchus cluster}
\label{model}

We have mentioned earlier in this paper that the radio flux density
values used previously by others to model the SED of the core of the
Ophiuchus cluster significantly overestimated the actual ones, due to
the huge synthesized beams of the old radio observations (Johnston et
al. 1981; see also Figure 1).  Therefore, the use of those radio flux
densities  to constrain the properties of the Ophiuchus
SED (Eckert et al. 2008) as well as to test proposed models for dark
matter annihilation (Profumo 2008) should be updated according to the
data presented earlier in this paper.  In this section, we model and
interpret the SED of the Ophiuchus cluster of galaxies, using our
high-resolution, high-sensitivity measurements at 240 and 607 MHz,
along with the 74 MHz VLA measurements (using archival data), the recent 
1400 MHz VLA measurements of the Ophiuchus radio mini-halo (Govoni et al. 2009,
Murgia et al. 2009), and the high-energy measurements quoted in Table 
\ref{all_data}.  In particular, and as shown in previous sections, we obtained 
upper limits to the diffuse non-thermal radio emission of the cluster at low
frequencies and an upper limit to luminosity of the non-thermal 
emission from \emph{XMM-Newton} data (see Sect.~\ref{xrays}).


\subsection{Dark matter models}

The new radio measurements permit to set stringent limits on the
synchrotron radiation produced by a population of non-thermal
electrons and positrons. In turn, this sets limits on weakly
interacting massive particle (WIMP) dark matter models. Indeed, the
pair-annihilation of WIMPs produces, as a result of the hadronization
and decay of particles in the standard model, energetic electrons and
positrons. The limits that can be set depend on the WIMP model
(specifically, on the WIMP mass, pair-annihilation cross section and
dominant annihilation final state), as well as on the dark matter
density profile and on the average magnetic field in the cluster. We
refer the reader to Totani (2004), Colafrancesco, Profumo \& Ullio (2006),
Colafrancesco, Profumo \& Ullio (2007), Jeltema and Profumo (2008) and Profumo
(2008), where the multi-frequency emission from WIMP annihilation has
been discussed in detail. In particular, we use here the same approach
as in Colafrancesco et al. (2006) and in Profumo (2008), and model the
dark matter density distribution in the Ophiuchus cluster according to
what described in Profumo (2008).

The new limits on the radio emission are compatible, in principle,
with the WIMP annihilation interpretation of the hard X-ray flux from
Ophiuchus presented in Profumo (2008). However, the magnetic field
required to be consistent with our new measurements is extremely low,
of the order of $10^{-2}$ to $10^{-3}$ $\mu$G.

Nonetheless, we can turn our radio data into limits on WIMP models.
For this purpose, we make more conservative assumptions on the density
profile and substructures than those made in Profumo (2008). In
particular, we assume a concentration $c_{\rm vir}=10$ and a Navarro,
Frenk and White profile (Navarro, Frenk \& White 1996). In addition,
we assume that 10\% of the mass belongs to substructures, and that the
average concentration bias in subhalos versus the host halo is
$\langle c_s\rangle/\langle c_{\rm vir}\rangle\sim 2$, which gives a
weighted enhancement factor $\Delta^2\sim6\times 10^5$, in the
notation of Eq.(29-31) of Colafrancesco et al. (2006). Integrating
over the line of sight the number density of dark matter particle
pairs, we find that the radiation resulting from WIMP annihilations in
Ophiuchus is given by:

\begin{eqnarray}
\frac{{\rm d}\Phi_\gamma}{{\rm d}E_\gamma} & = &2.4\times 10^{-13}\left(\frac{\langle\sigma\ v\rangle_0}{3\times 10^{-26}\ {\rm cm}^{3}{\rm s}^{-1}}\right) \nonumber \\
& \times & \left(\frac{100\ {\rm GeV}}{m_{\rm WIMP}}\right)^2\ \frac{{\rm d}N_\gamma}{{\rm d}E_\gamma}\ {\rm GeV}^{-1}{\rm cm}^{-2}{\rm s}^{-1}
\end{eqnarray}

where ${{\rm d}N_\gamma}/{{\rm d}E_\gamma}$ indicates the flux of
photons resulting from one dark matter annihilation event. Had we
assumed a cored profile for the dark matter distribution, we would
have obtained a flux a factor 2 smaller than indicated in the previous
equation: unlike for nearby structures, the dependence on the details
of the dark matter density distribution are not clearly for distant
systems.

\begin{figure}
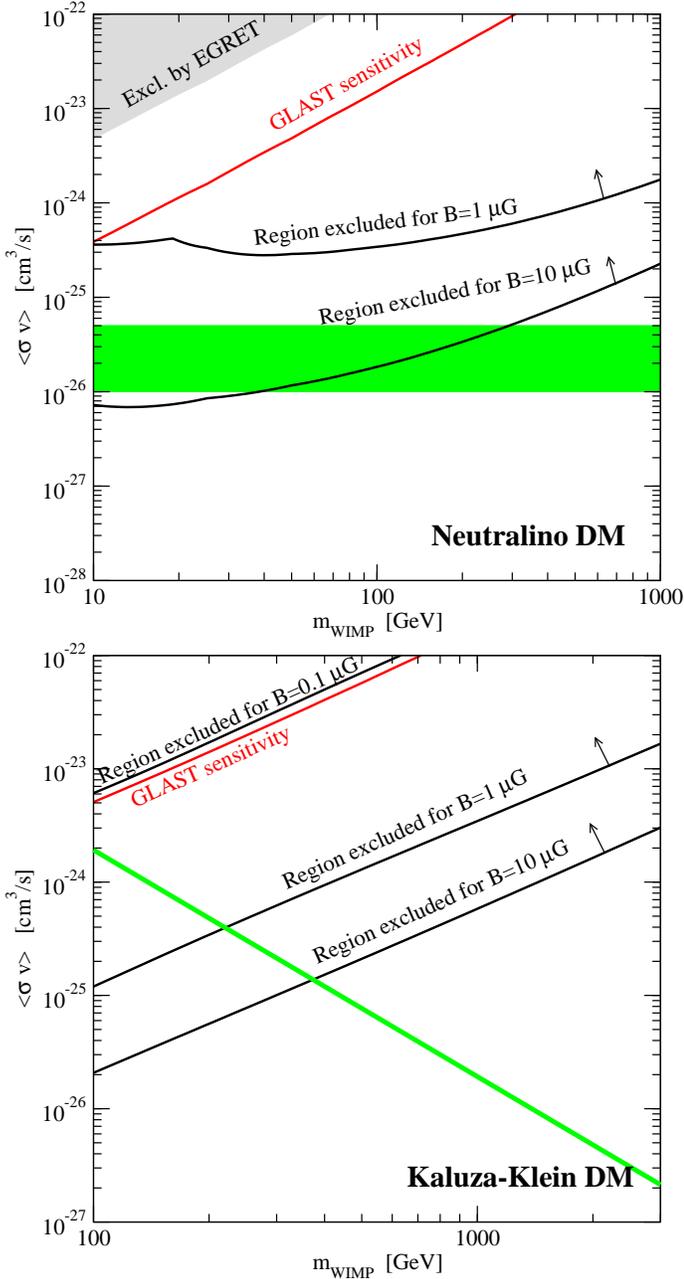

\mbox{\includegraphics[height=8.5cm]{DM_NEUT_DEF.eps}}
\mbox{\includegraphics[height=8.5cm]{DM_UED_DEF.eps}}
\caption{The plane of mass and pair-annihilation cross section for
  neutralino-like dark matter (top) and Kaluza-Klein dark matter
  (bottom). Regions above the black lines are ruled out by radio data,
  for the value of the magnetic field indicated on the curves. The
  Fermi-LAT sensitivity extends down to the red lines, while the gray
  region is excluded by EGRET. The green regions indicate the
  preferred portions of the parameter space in the two
  models.\label{fig:dm}}
\end{figure}

\begin{figure}
\begin{center}
\mbox{\includegraphics[height=8.5cm]{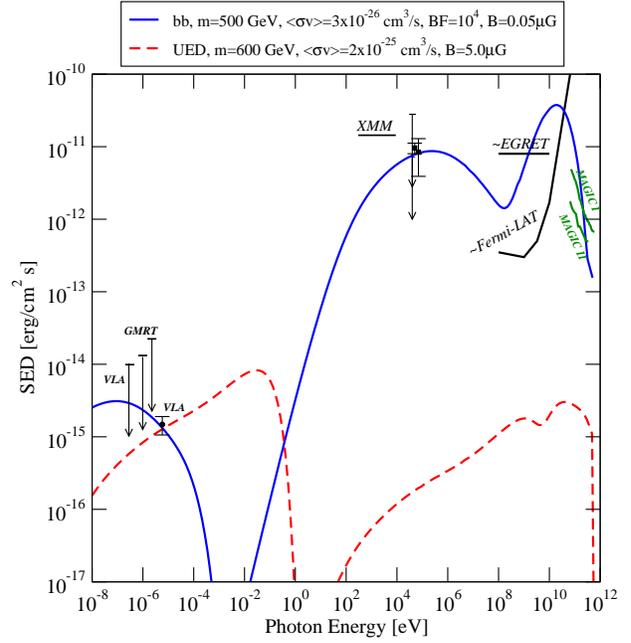}}
\caption{The SED for two dark matter models. Note that if the magnetic
  field is around the $\mu$G level, the limits we obtain are much
  better than the Fermi-LAT performance, and actually put rather
  stringent constraints on dark matter models. Only if the average
  magnetic field is extremely low (solid blue line), then Fermi-LAT
  measurements will be of very much use for constraining currently
  proposed models. In this and the following SED plots, the solid
  points in the X-ray band correspond to the INTEGRAL data, while the
  two arrows in that energy band correspond to the upper limits from
  Swift/BAT and Suzaku; the XMM and EGRET upper limits are also
  indicated.
\label{fig:dm_sed}}
\end{center}
\end{figure}

We show in the top panel of Fig.~\ref{fig:dm} the limits we obtain, at
the 95\% C.L., for a final state typical of neutralino dark matter in
the minimal supersymmetric extension of the standard model, namely
$b\bar b$ (Jungman, Kamionkowski \& Griest 1996, Bertone, Hooper \&
Silk 2005), for two values of the magnetic field. We also shade in
gray the region excluded by the EGRET limit on the gamma-ray flux
(integrated flux above 0.1 GeV larger than $5\times 10^{-8}$ ph/$({\rm
  cm}^2\ {\rm s})$, Reimer et al. 2003), and the line corresponding to
the sensitivity of the Large Area Telescope (LAT) on-board the Fermi
Gamma-Ray Space Telescope (formerly known as GLAST) sensitivity in one
year of operations. For the latter, we assume a sensitivity of
$2\times 10^{-10}$ ph/$({\rm cm}^2\ {\rm s})$ above 1
GeV\footnote{http://www-glast.slac.stanford.edu/software/IS/
  glast\_lat\_performance.htm}. The green band shows the range of
values for the pair annihilation cross section typically associated
with thermal dark matter production in the early universe (Jungman,
Kamionkowski \& Griest 1996, Bertone, Hooper \& Silk 2005). The bottom
panel indicates the limits we get for Kaluza-Klein WIMP dark matter in
the context of Universal Extra-Dimensions (Hooper \& Profumo
2007). The green line shows the typical value for the cross section
expected in those models (specifically in the minimal setup). We
notice that with a magnetic field of the order of 1\,$\mu$G, we are
able to place rather stringent limits on dark matter models. The
precise limits, however, depend crucially on the average magnetic
field.  Fig.~\ref{fig:dm_sed} illustrates that there exist models that
fit both the VLA radio data point and the hard X-ray emission from
Ophiuchus (blue solid line).  The implied needed boost factor from
dark matter substructures is however very large, and the magnetic
field is very small ($\sim$50 nG). Large boost factors can be due to
either a non-standard cosmological setup or a dark matter production
mechanism in the early universe, or to a more extreme substructure
setup than what employed in Profumo (2008), or to a combination of
both effects.  Clearly, for reasonable values of the magnetic field,
radio data put the most stringent limits on dark matter annihilation
in the Ophiuchus cluster. In Fig.~\ref{fig:dm_sed}, the MAGIC~I and
MAGIC~II Imaging Atmospheric Cherenkov Telescope (IACT) sensitivities
are also shown, as in the following SED plots, as reported in Carmona
et al. (2007).

Using a smaller value for the concentration, the number
density of particle dark matter pairs decreases, although slightly:
for instance, employing $c_{\rm vir}=6$ instead of 10 we get a
reduction of less than 10\% in the overall normalization factor.

\subsection{Synchrotron + Inverse Compton models}

We also considered Inverse Compton (IC) models, where the hard X-ray
emission is due to IC scattering of a population of non-thermal
electrons off the cosmic microwave background. We assume that the
electron population responsible for the hard X-ray emission has a
spectral distribution of the form:

\begin{equation}
\frac{{\rm d}N_e}{{\rm d}E_e}=N_0\left(\frac{E_e}{1\ {\rm GeV}}\right)^{-\alpha}\times \exp[-E_e/E_c],
\end{equation}

i.e. a power-law with index $\alpha$ plus an exponential cut-off at
electron energy equal to $E_c$.  The election of such a relativistic
electron distribution is justified, since radio mini-halos can be
explained in terms of primary relativistic electrons re-accelerated by
magneto-hydrodynamic turbulence (Ferrari et al. 2008). The importance
of this mechanism was recently confirmed by Brunetti et al. (2008),
who also noticed that spectrum cut-offs are a well-known signature of
the turbulence acceleration, implying a corresponding cut-off in the
emitting electron population.  We note that due to the exponential
cut-off in the electron population, there is not an analytical
solution that directly links the index $\alpha$ of the electron
population with the observed radio spectral index discussed in section
\ref{sources}.  We set the normalization $N_0$ to obtain the best
possible fit to the \emph{INTEGRAL} hard X-ray data (Eckert et
al. 2008). We then compute the integrated gamma-ray flux and the
maximal value for the magnetic field compatible at the 95\% C.L. along
with our upper limits of the diffuse non-thermal radio emission from
the core of the Ophiuchus cluster.

\begin{figure}
\mbox{\includegraphics[height=8.5cm]{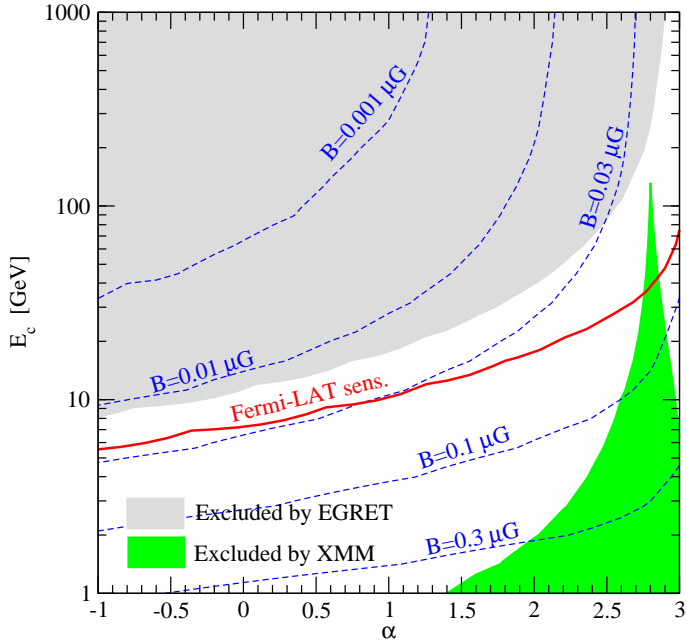}}
\mbox{\includegraphics[height=8.5cm]{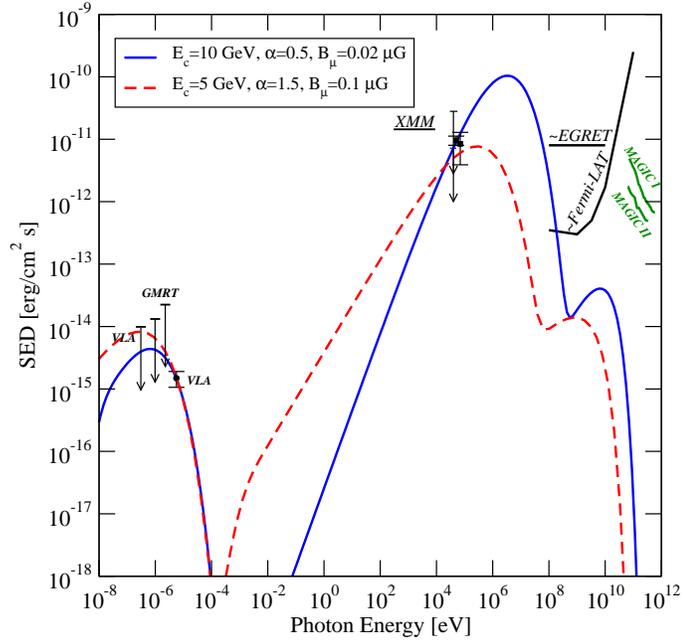}}
\caption{Top: The ($E_c,\alpha$) plane for models fitting the hard
  X-ray data. The pairs of $\alpha,E_c$ excluded by soft X-ray and
  gamma-ray data are shaded in green and gray, respectively.  the
  Fermi-LAT sensitivity is indicated with a red line.  We also show
  the values of the magnetic field required to have a radio emission
  compatible with the observations. Bottom: SED for two models
  compatible with the radio and non-thermal hard-X-ray data. (See main
  text for details.)
  \label{fig:ichxr}}
\end{figure}

The top panel of Fig.  \ref{fig:ichxr} shows the parameter space for
the cut-off electron energy, $E_c$, versus the injected index,
$\alpha$, of the relativistic electron population, which are the only
two free parameters in Equation 2. Shown are also the lines of
constant magnetic field, As it is evident from this graphic, very low
($\lsim$0.015$\mu$G) values of the magnetic field are excluded by the
EGRET data. Similarly, values significantly higher than
$\sim$0.3$\mu$G can be mostly excluded because they enter into
conflict with the \emph{XMM}-Newton data, or require unreasonable
high, or low, values of $\alpha$. The preferred values for the average
magnetic is in the range from $\sim$0.02$\mu$G up to $\sim$0.3$\mu$G.
The bottom panel of Fig. \ref{fig:ichxr} shows the resulting SED for
two representative models, which are compatible with the new radio
data, the nonthermal X-ray INTEGRAL detection and our upper limit to
the non-thermal X-ray emission from \emph{XMM}-Newton. The second peak
at very high energies ($\gsim$ 1 GeV) corresponds to the non-thermal
Bremsstrahlung emission.  This plot also shows that future
high-sensitivity, high-angular resolution radio data, along with
sensitive measurements at high energies will be of much use to
further constrain the SED of Ophiuchus within this scenario.

\subsection{PeV models}

Another plausible scenario to account for 
both the radio and hard X-ray data is that outlined in Inoue, Aharonian
\& Sugiyama (2005), who find that very high energy protons resulting from strong
accretion shocks produce high energy (TeV-PeV) electron-positron pairs
via interactions with the cosmic microwave background, which then
radiate synchrotron and inverse Compton emission, peaking respectively
at hard X-ray and TeV gamma-ray energies. For simplicity, we modelled
the injected electron-positron spectrum as mono-energetic, at an
energy specified by the parameter $E_{\rm inj}$. We then computed the
equilibrium spectrum by accounting for IC, synchrotron, Bremsstrahlung
and Compton energy losses, and neglected spatial diffusion. The
resulting electron-positron equilibrium spectrum shows a spectral
index $\alpha\sim 2$ at large energies (between where IC and
Synchrotron energy losses dominate, $E\sim1$ GeV, and the cutoff
energy $E_c$), and a sharp cut-off at $E_c=E_{\rm inj}$. We then
fitted for the hard X-ray data and computed the emission at all other
frequencies after having re-normalized the electron-positron flux
appropriately.

\begin{figure}
\mbox{\includegraphics[height=8.5cm]{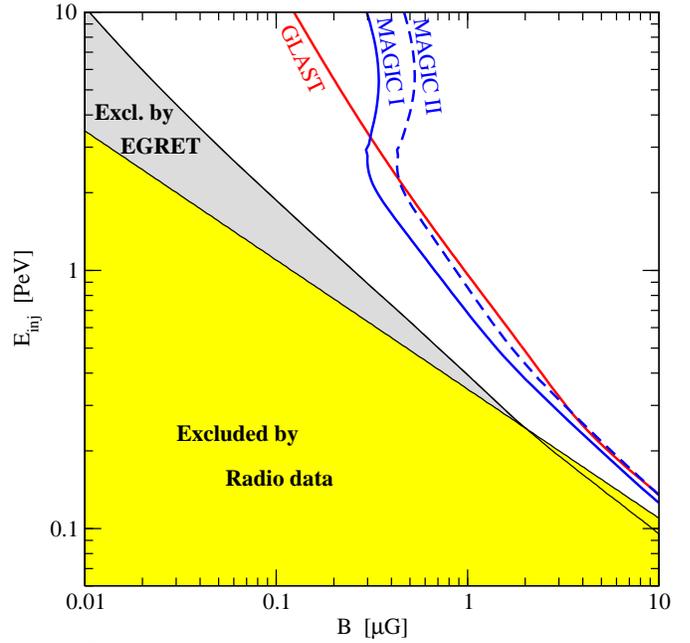}}
\mbox{\includegraphics[height=8.5cm]{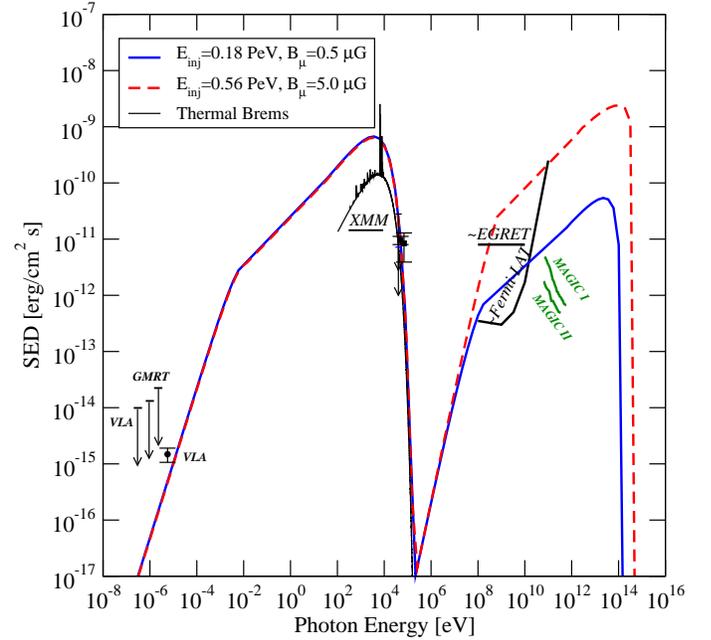}}
\caption{Top: The ($E_{\rm inj},B$) plane for PeV models fitting the
  hard X-ray data. The region shaded in yellow is ruled out by radio
  data, while the region shaded in gray is ruled out by the EGRET
  upper limit. Bottom: SED for the two models compatible with the hard
  X.ray data. Note that the models predict a much larger excess of
  X-rays with respect to the thermal Bremsstrahlung measured by
  \emph{INTEGRAL}, which rules them out.
  \label{fig:pev}}
\end{figure}

We studied the $(B,E_{\rm inj})$ plane (top panel of
Fig.~\ref{fig:pev}). The yellow area is excluded by the radio data,
while the gray area is ruled out by the EGRET constraints on the
gamma-ray emission from the Ophiuchus cluster. We also show in the
bottom panel the SED of two models in agreement with the \emph{INTEGRAL}
non-thermal detection.
Note that both models give comparable synchrotron fluxes (from radio
to X-rays), due to a degeneracy of the $E_{\rm inj}$ and $B$
parameters, as it is apparent in the top panel of
Fig. \ref{fig:pev}. The IC peak, instead, depends on the normalization
of the electron-positron flux, and this in turn depends only on the
magnetic field. Therefore, larger magnetic fields imply a suppressed
IC emission.  As the red dashed model shows, IC models with a soft
injection spectral index ($\alpha \gsim 1.5$), and that explain both
radio and HXR data are generically in tension with the EGRET
constraint on the gamma-ray flux.  The high energy tail of the
synchrotron emission is at a level compatible with the non-thermal
hard X-ray data reported by \emph{INTEGRAL}, but is not compatible
with the upper limit to the non-thermal emission in the soft X-ray
band that we measured from the \emph{XMM}-Newton data.  In addition,
the non-thermal soft X-ray emission predicted by the PeV models
largely exceeds the thermal Bremsstrahlung measured by
\emph{INTEGRAL}, and therefore this family of models can be ruled out.

\section{Summary}
\label{summary}

We present high resolution 240 and 607 MHz GMRT radio observations,
complemented with 74 MHz archival VLA radio observations of the
Ophiuchus cluster of galaxies, whose mini-radio halo has been recently
detected at 1400 MHz. We also present archival \emph{Chandra} and
\emph{XMM-Newton} data of the Ophiuchus cluster.  Our observations do
not show significant radio emission from the mini-halo, hence we
present upper limits to the integrated, diffuse non-thermal radio
emission of the core of the Ophiuchus cluster.

The \emph{XMM-Newton} observations can be well explained by a
two-temperature thermal model with temperatures of $\simeq$1.8 keV and
$\simeq$9.0 keV, respectively. This result is in agreement with
previous ones by Fujita et al. (2008), which suggested that the
central region of the Ophiuchus cluster is a cooling core.  In turn,
this result is consistent with the occurrence of a mini-halo, as
expected to be found in hot clusters with cool cores, and therefore
support the conclusion by Fujita et al. (2008) that Ophiuchus is not
the result of a major merger.  We also used the \emph{XMM-Newton} data
to set up an upper limit to the (non-thermal) X-ray emission from the
cluster and therefore constrain the SED of the Ophiuchus cluster.  

The high spatial resolution of the radio and X-ray data have allowed
us to study separately the diffuse emission of the cluster core from
that due to the population of point sources. In particular, we have
found a relatively bright radio and X-ray (\emph{Chandra}) point
source at the center of the cluster (source A in our images), which is
spatially coincident with the \emph{INTEGRAL} hard X-ray source. The
flat radio spectrum of this point source and the lack of non-thermal
emission has allowed us to associate it to a low-level AGN activity, most
likely associated to the cD central galaxy 2MASX
J17122774-2322108. This central source is very unlikely to be
responsible for the hard X-ray emission detected by \emph{INTEGRAL},
in spite of its apparent spatial coincidence.

Neither our newly obtained radio data at 240 and 607 MHz, nor the
archival VLA data at 74 MHz for the Ophiuchus cluster of galaxies
detect significant diffuse radio emission from the mini-halo in
Ophiuchus, which has been recently at 1400 MHz.  The radio and X-ray
non-thermal data have dramatic implications for currently proposed
scenarios to model the spectral energy distribution (SED) of the
Ophiuchus cluster of galaxies.  In particular, a synchrotron+IC model
is in agreement with the currently available data, if the average
magnetic field is between $\sim$0.02$\mu$G and $\sim$0.3$\mu$G.  A
pure dark matter scenario can also, in principle, reproduce both radio
and X-ray emission, but very large boost factors from dark matter
substructures and extremely low values of the average magnetic field
are needed.  We notice that for a magnetic field value comparable with
the expectations from Faraday rotation measures ($\sim 1 \mu$G),
GLAST-Fermi and high-resolution radio searches for a diffuse signal
from galaxy clusters will have a comparable and complementary reach in
the particle dark matter parameter space.  Finally, we also explored
the possibility of synchrotron and inverse Compton emission arising
from PeV electron-positron pairs (via interactions with the CMB), as
described in Inoue, Aharonian \& Sugiyama (2005). Those models can be
ruled out, since they predict a non-thermal soft X-ray emission that
largely exceeds the thermal Bremsstrahlung measured by \emph{INTEGRAL}
and the non-thermal X-ray emission inferred from the \emph{XMM}-Newton
data.

Given that Ophiuchus is one of the closest galaxy clusters, it can be
studied at high resolution and with high sensitivity. Therefore, its
study at all wavelengths is encouraged, since it can result in much
better constraints of its SED, and will allow for tests on acceleration
scenarios, as we have discussed in this paper. In particular, optical
spectroscopy is advisable, as it would allow to confirm, or rule out,
the membership of some of the sources shown in the radio images
presented in this work. Gamma-ray observations would also be very
useful to constrain the SED and the particle acceleration mechanisms.


\section*{Acknowledgments}
We are grateful to the referee for useful comments and suggestions in
the paper, and to Jukka Nevalainen for comments, suggestions and
helpful advice on the reduction of the \emph{XMM}-Newton data. We also thank
Valent\'i Bosch-Ramon for useful advices.  GMRT is run by the National
Centre for Radio Astrophysics of the Tata Institute of Fundamental
Research.  NRAO is a facility of the National Science Foundation
operated under cooperative agreement by Associated Universities,
Inc. This research has made extensive use of NASA's ADS, and of the
NASA/IPAC Extragalactic Database (NED).  M.A. P\'erez-Torres is a
Ram\'on y Cajal Post Doctoral Research Fellow funded by the Spanish
Ministry of Education and Science (MEC) and the Spanish Research
Council (CSIC).  F. Zandanel acknowledges the support of a JAE
PreDoc-CSIC fellowship in the IAA-CSIC, in Granada.
M.A.P\'erez-Torres, F. Zandanel and F. Prada also acknowledge support
by the Spanish MEC through grants AYA 2006-14986-C02-01 and AYA
2005-07789.  S. Profumo is partly supported by US Department of Energy
Contract DE-FG02-04ER41268 and by NASA Grant Number NNX08AV72G.
F. Panessa acknowledges support by ASI I/008/07 grants.


\label{lastpage}

\begin{thebibliography}{}

\bibitem[\protect\citeauthoryear{Ajello et al.}{2008}]{2008arXiv0809.0006A} 
Ajello M., et al., 2008, arXiv, arXiv:0809.0006

\bibitem[\protect\citeauthoryear{Atoyan {\ 
V&ouml}lk}{2000}]{2000ApJ...535...45A} Atoyan A.~M., V{\"o}lk H.~J., 2000, ApJ, 535, 45 

\bibitem[\protect\citeauthoryear{Baars et 
al.}{1977}]{1977A&A....61...99B} Baars J.~W.~M., Genzel R., Pauliny-Toth I.~I.~K., Witzel A., 1977, A\&A, 61, 99 

\bibitem[\protect\citeauthoryear{Bertone, Hooper, 
\& Silk}{2005}]{2005PhR...405..279B} Bertone G., Hooper D., Silk J., 2005, PhR, 405, 279

\bibitem[\protect\citeauthoryear{Blasi, Gabici, 
\& Brunetti}{2007}]{2007astro.ph..1545B} Blasi P., Gabici S., Brunetti G., 2007, astro, arXiv:astro-ph/0701545

\bibitem[\protect\citeauthoryear{Brunetti et 
al.}{2008}]{2008Natur.455..944B} Brunetti G., et al., 2008, Natur, 455, 944

\bibitem[\protect\citeauthoryear{Carmona et 
al.}{2007}]{2007arXiv0709.2959C} Carmona E., et al., 2007, arXiv, arXiv:0709.2959 

\bibitem[\protect\citeauthoryear{Cohen et al.}{2007}]{2007AJ....134.1245C} 
Cohen A.~S., Lane W.~M., Cotton W.~D., Kassim N.~E., Lazio T.~J.~W., Perley 
R.~A., Condon J.~J., Erickson W.~C., 2007, AJ, 134, 1245 

\bibitem[\protect\citeauthoryear{Colafrancesco, Profumo, 
\& Ullio}{2006}]{2006A&A...455...21C} Colafrancesco S., Profumo S., Ullio P., 2006, A\&A, 455, 21 

\bibitem[\protect\citeauthoryear{Colafrancesco, Profumo, 
\& Ullio}{2007}]{2007PhRvD..75b3513C} Colafrancesco S., Profumo S., Ullio P., 2007, PhRvD, 75, 023513 

\bibitem[\protect\citeauthoryear{Condon et al.}{1998}]{1998AJ....115.1693C} 
Condon J.~J., Cotton W.~D., Greisen E.~W., Yin Q.~F., Perley R.~A., Taylor 
G.~B., Broderick J.~J., 1998, AJ, 115, 1693 

\bibitem[\protect\citeauthoryear{Dunn 
\& Fabian}{2006}]{2006MNRAS.373..959D} Dunn R.~J.~H., Fabian A.~C., 2006, MNRAS, 373, 959 

\bibitem[\protect\citeauthoryear{Eckert et 
al.}{2008}]{2008A&A...479...27E} Eckert D., Produit N., Paltani S., Neronov A., Courvoisier T.~J.-L., 2008, A\&A, 479, 27 

\bibitem[\protect\citeauthoryear{Ensslin 
\& Biermann}{1998}]{1998A&A...330...90E} Ensslin T.~A., Biermann P.~L., 1998, A\&A, 330, 90 

\bibitem[\protect\citeauthoryear{Feretti 
\& Giovannini}{2007}]{2007astro.ph..3494F} Feretti L., Giovannini G., 2007, astro, arXiv:astro-ph/0703494 

\bibitem[\protect\citeauthoryear{Ferrari et 
al.}{2008}]{2008SSRv..134...93F} Ferrari C., Govoni F., Schindler S., Bykov 
A.~M., Rephaeli Y., 2008, SSRv, 134, 93 

\bibitem[\protect\citeauthoryear{Fujita et al.}{2008}]{2008PASJ...60.1133F} 
Fujita Y., et al., 2008, PASJ, 60, 1133

\bibitem[\protect\citeauthoryear{Govoni et al.}{2009}]{2009arXiv0901.1941G} 
Govoni F., Murgia M., Markevitch M., Feretti L., Giovannini G., Taylor 
G.~B., Carretti E., 2009, arXiv, arXiv:0901.1941 

\bibitem[\protect\citeauthoryear{Hasegawa et 
al.}{2000}]{2000MNRAS.316..326H} Hasegawa T., et al., 2000, MNRAS, 316, 326 

\bibitem[\protect\citeauthoryear{Hooper 
\& Profumo}{2007}]{2007PhR...453...29H} Hooper D., Profumo S., 2007, PhR, 453, 29 

\bibitem[\protect\citeauthoryear{Inoue, Aharonian, 
\& Sugiyama}{2005}]{2005ApJ...628L...9I} Inoue S., Aharonian F.~A., Sugiyama N., 2005, ApJ, 628, L9 

\bibitem[\protect\citeauthoryear{Jeltema 
\& Profumo}{2008}]{2008ApJ...686.1045J} Jeltema T.~E., Profumo S., 2008, ApJ, 686, 1045 

\bibitem[\protect\citeauthoryear{Johnston et 
al.}{1981}]{1981ApJ...245..799J} Johnston M.~D., Bradt H.~V., Doxsey R.~E., 
Marshall F.~E., Schwartz D.~A., Margon B., 1981, ApJ, 245, 799

\bibitem[\protect\citeauthoryear{Jungman, Kamionkowski, 
\& Griest}{1996}]{1996PhR...267..195J} Jungman G., Kamionkowski M., Griest K., 1996, PhR, 267, 195 

\bibitem[\protect\citeauthoryear{Molendi 
\& Gastaldello}{2008}]{2008arXiv0807.2653M} Molendi S., Gastaldello F., 2008, arXiv, arXiv:0807.2653 

\bibitem[\protect\citeauthoryear{Murgia et al.}{2009}]{2009arXiv0901.1943M} 
Murgia M., Govoni F., Markevitch M., Feretti L., Giovannini G., Taylor 
G.~B., Carretti E., 2009, arXiv, arXiv:0901.1943 

\bibitem[\protect\citeauthoryear{Navarro, Frenk, 
\& White}{1996}]{1996ApJ...462..563N} Navarro J.~F., Frenk C.~S., White S.~D.~M., 1996, ApJ, 462, 563

\bibitem[\protect\citeauthoryear{Profumo}{2008}]{2008PhRvD..77j3510P} 
Profumo S., 2008, PhRvD, 77, 103510 

\bibitem[\protect\citeauthoryear{Renaud et 
al.}{2006}]{2006A&A...453L...5R} Renaud M., B{\'e}langer G., Paul J., Lebrun F., Terrier R., 2006, A\&A, 453, L5

\bibitem[\protect\citeauthoryear{Reimer et al.}{2003}]{2003ApJ...588..155R} 
Reimer O., Pohl M., Sreekumar P., Mattox J.~R., 2003, ApJ, 588, 155

\bibitem[\protect\citeauthoryear{Sanders et 
al.}{2004}]{2004MNRAS.349..952S} Sanders J.~S., Fabian A.~C., Allen S.~W., 
Schmidt R.~W., 2004, MNRAS, 349, 952 

\bibitem[\protect\citeauthoryear{Sanders, Fabian, 
\& Dunn}{2005}]{2005MNRAS.360..133S} Sanders J.~S., Fabian A.~C., Dunn R.~J.~H., 2005, MNRAS, 360, 133 

\bibitem[\protect\citeauthoryear{Sarazin}{1999}]{1999astro.ph.11439S} 
Sarazin C.~L., 1999a, astro, arXiv:astro-ph/9911439

\bibitem[\protect\citeauthoryear{Sarazin}{1999}]{1999ApJ...520..529S} 
Sarazin C.~L., 1999b, ApJ, 520, 529 

\bibitem[\protect\citeauthoryear{Timokhin, Aharonian, 
\& Neronov}{2004}]{2004A&A...417..391T} Timokhin A.~N., Aharonian F.~A., Neronov A.~Y., 2004, A\&A, 417, 391

\bibitem[\protect\citeauthoryear{Watanabe et 
al.}{2001}]{2001PASJ...53..605W} Watanabe M., Yamashita K., Furuzawa A., 
Kunieda H., Tawara Y., 2001, PASJ, 53, 605 

\end{thebibliography}
\end{document}